\documentclass[a4paper]{PoS}
\usepackage{multirow}
\usepackage{color}

\title{What is the Astrophysical Meaning of the Intermediate Subgroup of GRBs?}

\ShortTitle{What is the Astrophysical Meaning of the Intermediate Subgroup of GRBs?}

\author{\speaker{Jakub \v{R}\'{\i}pa}%
        \thanks{We gratefully appreciate useful comments from Stephen Appleby.
        This study was supported by the Grant Agency of the Czech Republic - 
        Grant No. P209/10/0734, and by the Creative Research Initiatives Program (RCMST) of MSIP/NRF in Korea.}\\
        Institute of Basic Science, Sungkyunkwan University\\
        2066 Seobu-ro, Suwon, 440-746, Korea\\
        E-mail: \email{ripa.jakub@gmail.com}}

\author{Attila M\'esz\'aros\\
Charles University in Prague, Faculty of Mathematics and Physics, Astronomical Institute\\
V Hole\v{s}ovi\v{c}k\'ach 2, CZ 180 00 Prague 8, Czech Republic\\
E-mail: \email{meszaros@cesnet.cz}
}

\abstract{Published articles concerning the intermediate (third) subgroup of GRBs are surveyed.
From a statistical perspective this subgroup may exist, however its significance depends on which data set is used.
Its astrophysical meaning is unclear because the occurrence of this subgroup can also be an artificial selection effect.
Hence, GRBs from this subgroup need not be given by a physically different phenomenon. 
The aim of this contribution is to search for the answer to the question in the title.}

\FullConference{Swift: 10 Years of Discovery\\
		 2-5 December 2014\\
		 La Sapienza University, Rome, Italy}

\begin{document}

\begin{center}
\begin{figure}
\centering
\includegraphics[width=0.7\textwidth]{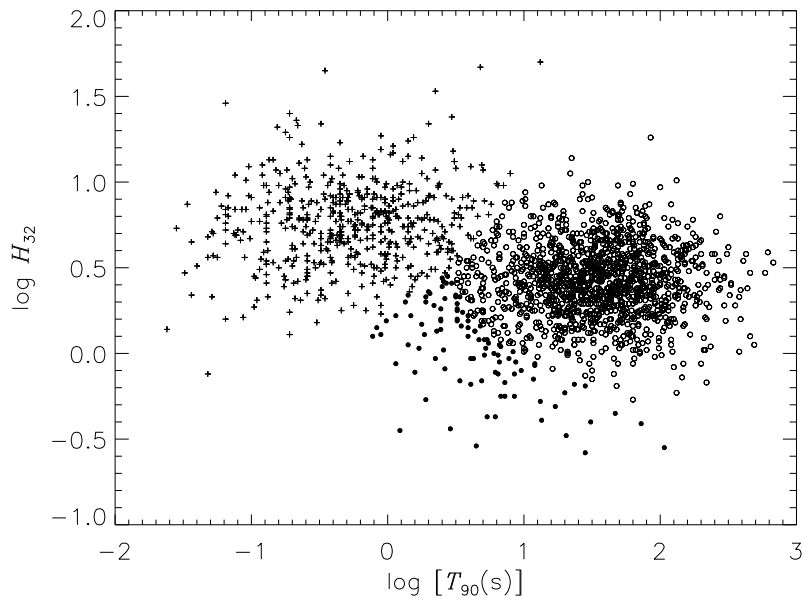}\\
\caption{Hardness ratio $H_{32}$ vs. $T_{90}$ duration of GRBs detected by CGRO-BATSE with identified 
groups of short (crosses), intermediate (full circles), and long (opened circles) bursts as published in \cite{hor06}.}
\label{fig:batse_H_T90}
\end{figure}
\end{center}

\section{Two - Three Different Groups of GRBs}

Gamma-Ray Bursts (GRBs) are fascinating cosmological objects, but they are not all of the same kind.  
There are at least two different groups, `short/hard' and `long/soft' \cite{maz81,kou93,nor01,bal03,bor04,me06b,zha09,brom13}.
A possibility of the existence of further groups has been intensively studied using various statistical techniques 
\cite{ver10,rip12,hor98,muk98,bal01,hor02,hor08,hor09,huj09,rip09,hor10,hor06,chat07,ugar11}.
It has been postulated that there might be a third group of GRBs with intermediate durations. However, statistical tests applied to
different datasets obtained from different satellites assign varying significance to this result. 
The astrophysical origin of this subgroup also remains unclear.

The three groups of GRBs found by BATSE, an instrument on board the Compton Gamma-Ray Observatory (CGRO),
are shown in Figure~\ref{fig:batse_H_T90}. The figure compares the durations $T_{90}$ and hardnesses $H_{32}$, 
i.e. the ratios of the received energy per unit area in the range $100-300$\,keV over the same quantity in the range $50-100$\,keV.
In the figure 1956 bursts are shown - they were observed by this instrument over the years 1991-2000.

The short/hard and the long/soft groups are clearly separated around $T_{90} \simeq 2$\,s.
It is now generally accepted \cite{bal03} that they are distinct astrophysical phenomena. The long ones are believed to be
coupled with supernovae type Ic. The physics of the short bursts remains unclear, although a merging of two compact 
objects such as neutron stars has been suggested \cite{gehl12}.

\begin{center}
\begin{figure}
\centering
\includegraphics[width=0.7\textwidth]{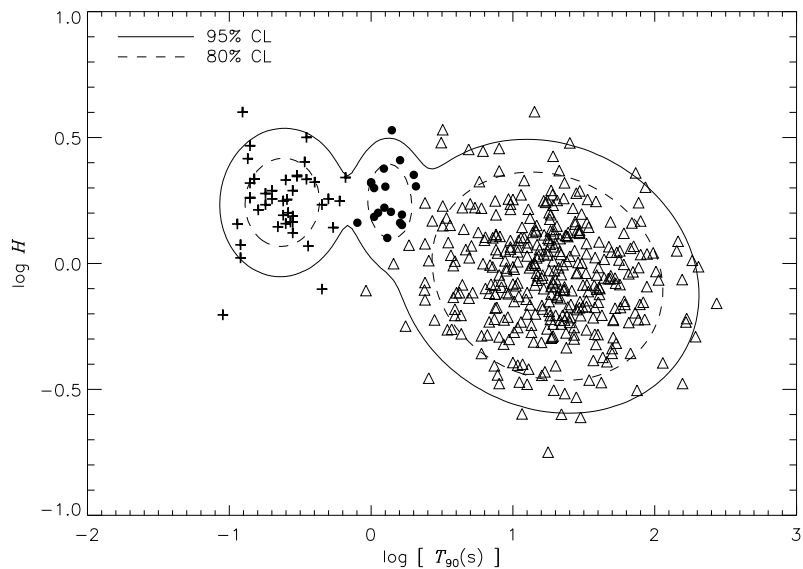}\\
\caption{Hardness ratio $H$ vs. duration $T_{90}$ of GRBs detected by RHESSI 
with identified groups of short (crosses), intermediate (full circles), and long (triangles) bursts as published in \cite{rip12}.}
\label{fig:rhessi_H_T90}
\end{figure}
\end{center}

\begin{center}
\begin{figure}
\centering
\includegraphics[width=0.64\textwidth]{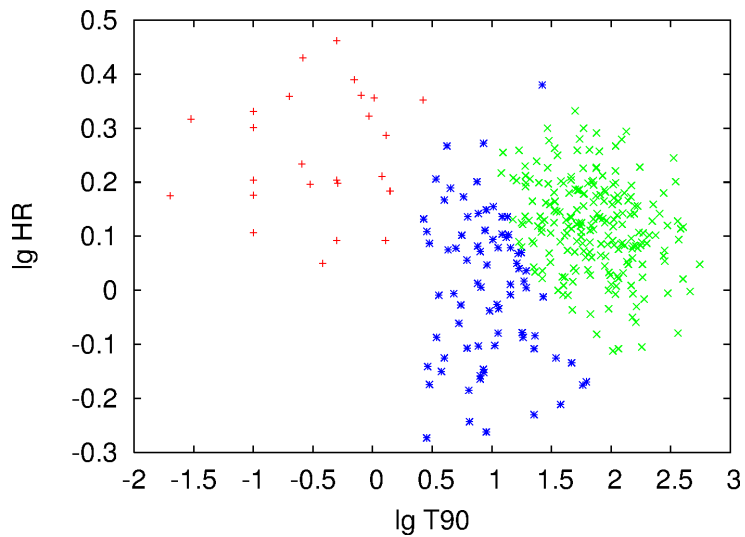}\\
\caption{Hardness ratio vs. duration $T_{90}$ of GRBs detected by Swift-BAT with identified 
groups of short (red pluses), intermediate (blue stars), and long (green $\times$) bursts as published in \cite{hor10}.}
\label{fig:swift_H_T90}
\end{figure}
\end{center}

Several statistical analyses show that the existence of an intermediate subclass cannot be excluded. Three distinct groups have been found -
not only in the BATSE\footnote{http://gammaray.msfc.nasa.gov/batse/} database, but also for the 
RHESSI\footnote{http://hesperia.gsfc.nasa.gov/hessi/index.html} (Figure~\ref{fig:rhessi_H_T90}) 
and Swift-BAT\footnote{http://swift.gsfc.nasa.gov/docs/swift/swiftsc.html} (Figure~\ref{fig:swift_H_T90}) 
databases (see \cite{rip12} and references therein).
Hence, from a statistical perspective, the existence of three subgroups is likely. 
However, it does not immediately follow that the three different subgroups arise from three astrophysically different progenitors.
There are several selection and instrumental biases \cite{hak00} which can cause these separations instead.

\section{The Physics of the Intermediate GRBs}

A key step in understanding the physics of the intermediate subgroup was made in \cite{ver10}. It was shown that
for the Swift database the intermediate subclass was related to X-Ray Flashes (XRFs). 
Since XRFs are related to the standard long/soft type GRBs \cite{kip03,sak05}, at least in the Swift database
the intermediate subgroup could simply be the tail of the long GRB distribution.

On the other hand, the GRBs of the RHESSI database's intermediate subgroup are not as soft as the long ones and
they do not appear to constitute a tail of the long GRBs (see Figure~\ref{fig:rhessi_H_T90}). In fact quite conversely,
they are more similar to the short bursts.
A detailed statistical analysis of the RHESSI database has shown that the intermediate group
in this database was similar to the short one \cite{rip12}.

\begin{center}
\begin{table}
\caption{Summary of published results concerning the GRB subgroups.
Mentioned are the significances of the third group found by different methods and using data from different instruments. 
F-test compares the best $\chi^2$ fits (two and three Gaussian distributions) of the $\log T_{90}$ duration. 
'ML r.' is the Maximum Likelihood ratio test applied either on the $\log T_{90}$ durations 
or on the \{$\log T_{90}$, $\log H$\} \{duration, hardness ratio\} pairs. 
'BIC' is the test based on the difference of the Bayesian Information Criterion values of the best 
fitted multivariate Gaussian components. BIC was also applied on the peak count rates $F$.}
\centering
\begin{tabular}{ccccc}
\\
\hline
Method           			    & CGRO-BATSE           			& Swift-BAT            			& RHESSI            				& BeppoSAX		\\
\hline
F-test \{$T_{90}$\}			    & $< 0.01$\,\% \cite{hor98}     		& 3.6\,\% \cite{huj09}  		& 6.9\,\% \cite{rip09}         			&         		\\
ML r. \{$T_{90}$\}	    		    & $0.5$\,\% \cite{hor02}     		& 0.5\,\% \cite{hor08}			& 0.04\,\% \cite{rip09} 			&  3.7\,\% \cite{hor09} \\
ML r. \{$T_{90}$, $H$\} 		    & $\lesssim 10^{-8}$\,\% \cite{hor06} 	& $\approx10^{-6}$\,\% \cite{hor10}	& 0.1\% \cite{rip09}, 0.3\% \cite{rip12} 	&         		\\
\multirow{ 2}{*}{BIC \{$T_{90}$, $H$\}}     &  						& 3 groups \cite{ver10}			& 2 groups \cite{rip12}				&         		\\
					    &  						& \footnotesize{(very strong evid.)}	& \footnotesize{(very strong evidence)}		&         		\\
\multirow{ 2}{*}{BIC \{$T_{90}$, $H$, $F$\}}&  						& 					& 3 groups \cite{rip12}				&         		\\
					    &  						& 					& \footnotesize{(very strong evidence)}		&         		\\
\multirow{ 2}{*}{Other methods}		    & $< 0.01$\,\% \cite{muk98} 		&                    			&           					&              		\\
					    &   3 groups \cite{bal01}, \cite{chat07}	&         				&               				&         		\\
\hline
\end{tabular}
\end{table}
\end{center}

\begin{center}
\begin{figure}
\centering
\includegraphics[width=0.64\textwidth]{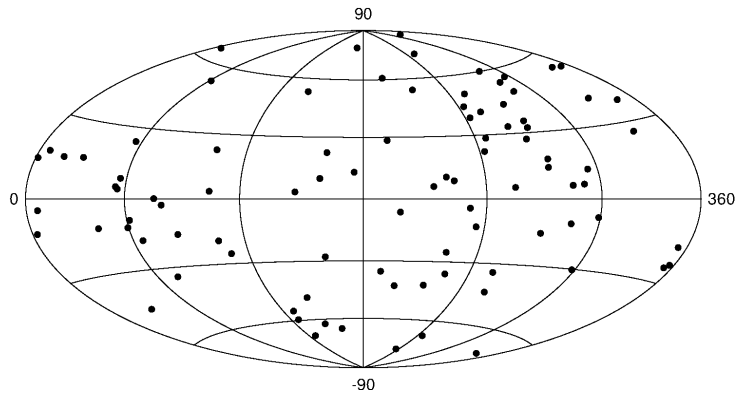}\\
\caption{Anisotropic distribution of 92 dim intermediate GRBs in equatorial coordinates from the BATSE database as published in \cite{mesz00}.}
\label{fig:batse_inter_anisot}
\end{figure}
\end{center}

For the BATSE database the physics of the intermediate GRBs remains an open question.
In addition, here a further interesting property exists.
The expected angular distribution of GRBs should be isotropic - this follows from the cosmological principle.
For the long GRBs this expectation can be fulfilled, but not for the short ones (for details see work \cite{vav08} and references therein).
Also the intermediate subgroup is not distributed isotropically (see Figure~\ref{fig:batse_inter_anisot}) on the sky \cite{mesz00}.

\section{Conclusion}

The separation of GRBs to the short/hard and long/soft groups and the connection of the long/soft group to supernovae is widely accepted. 
On the other hand, both the physics of the short/hard GRBs and the existence of intermediate GRBs remain open questions.
For the Swift database the intermediate GRBs can be related to XRFs and hence to the long bursts, 
but this relationship does not follow in the RHESSI database. For the BATSE dataset the relation between XRFs and the intermediate 
subgroup is also unclear. We conclude that instrumental effects are important and the identification of the intermediate subgroup with XRFs remains to be proven.

\end{document}